\documentclass[conference,9pt]{IEEEtran}
\IEEEoverridecommandlockouts
\usepackage{amsmath,graphicx}

\usepackage[dvipsnames]{xcolor}

\usepackage{amssymb,amsfonts}
\usepackage{graphicx}
\usepackage{textcomp}
\usepackage{subfigure}
\usepackage{textcomp} 
\usepackage{siunitx}  
\usepackage{algorithm}
\usepackage{algpseudocode}
\usepackage{xpatch}
\usepackage{pgfplots}
\usepackage{pgfplotstable}
\usepackage{readarray}
\usepackage{siunitx}
\usepackage{bm}
\usepackage[nolist]{acronym}
\usepackage{bbm}
\usepackage{gensymb}
\usetikzlibrary{external}
\usetikzlibrary{spy}

\definecolor{TUMBeamerYellow}    {rgb} {1.000,0.706,0.000}    
\definecolor{TUMBeamerOrange}    {rgb} {1.000,0.502,0.000}    
\definecolor{TUMBeamerRed}       {rgb} {0.898,0.204,0.094}    
\definecolor{TUMBeamerDarkRed}   {rgb} {0.792,0.129,0.247}    
\definecolor{TUMBeamerBlue}      {rgb} {0.000,0.600,1.000}    
\definecolor{TUMBeamerLightBlue} {rgb} {0.255,0.745,1.000}    
\definecolor{TUMBeamerGreen}     {rgb} {0.569,0.675,0.420}    
\definecolor{TUMBeamerLightGreen}{rgb} {0.710,0.792,0.510}    

\DeclareMathOperator{\T}{T}
\DeclareMathOperator{\He}{H}

\DeclareMathOperator{\inv}{-1}

\DeclareMathOperator{\diag}{diag}

\DeclareMathOperator{\imag}{j\hspace*{-1.5pt}}

\newcommand{\eye}{\bm{\mathrm{I}}}
\newcommand{\bbthet}{\bm{\bar{\theta}}}
\newcommand{\bD}{\bm{D}}

\newcommand{\Her}{{\He}}
\newcommand{\td}{{\text{d}}}
\newcommand{\tc}{{\text{c}}}

\newcommand{\tdk}{{\text{d},k}}

\newcommand{\bC}{\bm{C}}
\newcommand{\bQ}{\bm{Q}}

\newcommand{\NB}{N_{\text{B}}}

\newcommand{\NRe}{N_{\text{R}}}

\newcommand{\tre}{{\text{r}}}

\newcommand{\trek}{{\text{r},k}}
\newcommand{\ts}{{\text{s}}}
\newcommand{\bH}{\bm{H}}

\newcommand{\bh}{\bm{h}}
\newcommand{\mThet}{\bm{\Theta}}
\newcommand{\bu}{\bm{u}}
\newcommand{\bw}{\bm{w}}
\newcommand{\bv}{\bm{v}}
\newcommand{\bthet}{\bm{\theta}}

\newcommand{\cmplx}[1]{\mathbb{C}^{#1}}
\newcommand{\norm}[1]{\|#1\|}
\newcommand{\abs}[1]{|#1|}

\newcommand{\expct}[1]{\mathbb{E}[#1]}

\newcommand{\gaussdist}[2]{\mathcal{N}_{\mathbb{C}}(#1,#2)}
\newcommand{\summe}[2]{\sum_{#1}^{#2}}

\begin{document}

\begin{acronym}
    \acro{AoD}{angle of departure}
    \acro{AoA}{angle of arrival}
    \acro{ULA}{uniform linear array}
    \acro{CSI}{channel state information}
    \acro{LOS}{line of sight}
    \acro{EVD}{eigenvalue decomposition}
    \acro{BS}{base station}
    \acro{MS}{mobile station}
    \acro{mmWave}{millimeter wave}
    \acro{DPC}{dirty paper coding}
    \acro{IRS}{intelligent reflecting surface}
    \acro{AWGN}{additive white gaussian noise}
    \acro{MIMO}{multiple-input multiple-output}
    \acro{UL}{uplink}
    \acro{DL}{downlink}
    \acro{OFDM}{orthogonal frequency-division multiplexing}
    \acro{TDD}{time-division duplex}
    \acro{LS}{least squares}
    \acro{MMSE}{minimum mean square error}
    \acro{SINR}{signal to interference plus noise ratio}
    \acro{OBP}{optimal bilinear precoder}
    \acro{LMMSE}{linear minimum mean square error}
    \acro{MRT}{maximum ratio transmitting}
    \acro{M-OBP}{multi-cell optimal bilinear precoder}
    \acro{S-OBP}{single-cell optimal bilinear precoder}
    \acro{SNR}{signal to noise ratio}
    \acro{THP}{Tomlinson-Harashima Precoding}
    \acro{dTHP}{distributed THP}
    \acro{cTHP}{centralized THP}
    \acro{RIS}{reconfigurable intelligent surface}
    \acro{SE}{spectral efficiency}
    \acro{MSE}{mean squared error}
    \acro{ASD}{angular standard deviation}
    \acro{ZF-THP}{zero-forcing THP}
\end{acronym}

\title{Nonlinear Precoding\\ in the RIS-Aided MIMO Broadcast Channel}

\author{\IEEEauthorblockN{Dominik Semmler, Michael Joham, and Wolfgang Utschick}
\IEEEauthorblockA{\textit{School of Computation, Information and Technology, Technical University of Munich, 80333 Munich, Germany} \\
email: \{dominik.semmler,joham,utschick\}@tum.de}
}

\maketitle

\begin{abstract}
    We propose to use \ac{THP} for the \ac{RIS}-aided \ac{MIMO} broadcast channel where we assume a \ac{LOS} connection between the \ac{BS} and the \ac{RIS}.
    In this scenario, nonlinear precoding, like \ac{THP} or \ac{DPC}, has certain advantages compared to linear precoding as it is more robust in case the \ac{BS}-\ac{RIS} channel is not orthogonal to the direct channel.
    Additionally, \ac{THP} and \ac{DPC} allow a simple phase shift optimization which is in strong contrast to linear precoding for which the solution is quite intricate.
    Besides being difficult to optimize, it can be shown that linear precoding has fundamental limitations for statistical and random phase shifts which do not hold for nonlinear precoding.
    Moreover, we show that the advantages of \ac{THP}/\ac{DPC} are especially pronounced for discrete phase shifts.
        \end{abstract}

\begin{figure}[b]
    \onecolumn
    \centering
    \scriptsize{This work has been submitted to the IEEE for possible publication. Copyright may be transferred without notice, after which this version may no longer be accessible.}
    \vspace{-1.3cm}
    \twocolumn
\end{figure}

\section{Introduction}
\label{sec:intro}
\Acp{RIS} are viewed as a key technology for future communications systems (see \cite{Power_Min_IRS}).
These surfaces consist of many passive elements which allow to reconfigure the channel resulting in an improved system performance.
While this technology gives new degrees of freedom for system optimization, the high number of reflecting elements also drastically increase the complexity.
The channel estimation as well as the joint optimization of the precoders and the reflecting elements in a downlink scenario pose two major challenges for the technology.
We consider perfect \ac{CSI} in this article and the main focus will be on the second point, where we give a new algorithm for the joint optimization of the precoders and the reflecting elements in order to maximize the sum \ac{SE}.
For sum \ac{SE} maximization under perfect \ac{CSI}, there is already significant literature available, see, e.g., \cite{WSR,WMMSEMIMO,MIMOP2PCap}. 
Additionally, we assume that the \ac{BS}-\ac{RIS} channel is dominated by a \ac{LOS} channel which is motivated by the fact that both the \ac{BS} and the \ac{RIS} are both deployed in considerable height and in \ac{LOS} conditions by purpose (see \cite{LOSAssump}).
This assumption has already been considered in the literature (see, e.g. \cite{MaxMinSINR,LOSAssumpTwo,Eigenvalues}).

Recently, it has been shown in \cite{LOSRISDPCLinear} that especially nonlinear precoding techniques such as, e.g., \ac{DPC}, are particularly suited for such a scenario.
Following \cite{LOSRISDPCLinear}, it has been pointed out that it is crucial for the performance of the system that the \ac{BS}-\ac{RIS} channel is orthogonal to the direct channels.
This condition is not met in a practical scenario and \ac{DPC} is considerably more robust in comparison to linear precoding if this condition is not completely fulfilled.
Furthermore, linear precoding suffers from fundamental limitations when considering random or statistical phase shifts (see \cite{LOSRISDPCLinear}).
As an example, in a rank improvement scenario under Rayleigh fading, the \ac{SE} of linear precoding for random phase shifts is upper bounded by a constant whereas for \ac{DPC} it is increasing monotonically with the number of reflecting elements.
Because of the above mentioned benefits of \ac{DPC}, we propose an efficent algorithm in this arcticle based on \ac{THP} (see \cite{THPOne,THPTwo}).
\ac{THP} is a vector perturbation technique (see \cite{VectorPerturbation}) with reduced complexity which we show to share the same benefits as \ac{DPC}.
There exist fundamental advantages over linear precoding that are especially pronounced for random or statistical phase shifts.
We focus on random phase shifts and instantanteous phase shifts in this article, whereas statistical phase shifts will be analyzed in future work.
Particularly, for discrete phase shifts, we observe that also for instantanous phase shifts, \ac{THP} has a significant advantage.
In comparison to \ac{DPC}, \ac{THP} is simple to implement.
However, it also comes with a certain performance loss.
Specifically, it suffers from a shaping loss and we analyze its performance within this article.
Our contributions are:
\begin{itemize}
    \item We derive the high-\ac{SNR} optimal solution for \ac{DPC} in case where there are more users than base station antennas.
    Additionally, a low-complexity phase shift optimization for \ac{DPC} is derived which can be shown to be optimal for specific scenarios.
    \item We propose \ac{THP} for the \ac{RIS}-aided scenario and give a low-complexity algorithm based on that method.
        In particular, we use \ac{dTHP} which appears to be better suited in comparison to \ac{cTHP} in a \ac{RIS}-aided scenario.
    \item We compare \ac{DPC}, linear precoding, and \ac{THP} (including the shaping loss) in this article and show that \ac{THP} has the same benefits as \ac{DPC} shown in \cite{LOSRISDPCLinear},
            specifically, we show discrete phase shifts to be an interesting aspect in favor of \ac{THP}/\ac{DPC}.
\end{itemize}
It has to be noted that throughout the article, we consider the conventional phase shift model which is the most popular model in the literature.
However, a more accurate model has been given in \cite{NewChannelModel}.
The algorithms and results in this article can be directly extended to the model in \cite{NewChannelModel} and, additionally, to mutual coupling by considering decoupling networks (see \cite{DecouplingNetwork}).
\section{System Model}
A scenario with one \ac{BS} serving $K$ single-antenna users is considered.
The system is supported by one \ac{RIS} with $\NRe$ reflecting elements.
Furthermore, we assume the \ac{BS}-\ac{RIS} channel to be rank-one, i.e., $\bH_\ts = \bm{a}\bm{b}^\Her$, where w.l.o.g. we assume $\norm{\bm{b}}_2=1$.
Hence, the channel of user $k$ reads as
\begin{equation}
    \bh_k^\Her = \bh^\Her_\tdk + \bh_\trek^\Her \mThet \bm{a}\bm{b}^\Her \; \in \cmplx{1 \times \NB}
\end{equation}
with $\bh_\tdk^\Her \in \cmplx{1 \times \NB}$ being the direct channel from the \ac{BS} to the $k$-th user,  $ \bh_\trek^\Her \in \cmplx{1 \times \NRe}$ being the reflecting channel from the \ac{RIS}
to user $k$, and $\mThet = \diag(\bthet) \in \cmplx{\NRe \times \NRe}$ with $\bthet \in \{\bm{z} \in \cmplx{\NRe}: \abs{z_n}=1,\; \forall n\}$ being the phase manipulation at the \ac{RIS}.
By defining $\bh_{\tc,k}^\Her = \bh_{\tre,k}^\Her \diag(\bm{a})$ we arrive at the equivalent expression
\begin{equation}
    \bh_k^\Her = \bh^\Her_\tdk + \bh_{\tc,k}^\Her \bthet \bm{b}^\Her \; \in \cmplx{1 \times \NB}.
\end{equation}
Stacking the user channels into the composite channel matrices $\bm{H}_{\text{d}} =[\bh_{\td,1},\dots,\bh_{\td,K}]^\Her$ and $\bm{H}_{\tc} =[\bh_{\tc,1},\dots,\bh_{\tc,K}]^\Her$, we obtain 
\begin{equation}
    \bm{H} =    \bm{H}_{\text{d}} + \bm{H}_{\tc} \bthet \bm{b}^\Her
                \in \cmplx{ K \times \NB}
\end{equation}
where $\bm{H}=[\bh_{1},\dots,\bh_{K}]^\Her$. Due to $\norm{\bm{b}}_2=1$ and by defining
\begin{align}
    \label{eq:CiDefinition}
\bm{C}  &= \bm{H}_{\text{d}}\bm{P}_{\bm{b}}^{\perp} \bm{H}_{\text{d}}^{\Her},\quad
\bD = \begin{bmatrix}
    \bm{H}_{\text{c}},\; \bm{H}_{\text{d}}\bm{b}
\end{bmatrix},
\quad  \bbthet = \begin{bmatrix}
    \bthet^\Her & 1
\end{bmatrix}^\Her
\end{align}
the Gram channel matrix can be written (similar to \cite{Eigenvalues} and \cite{LOSRISDPCLinear}) as 
\begin{equation}\label{eq:RankOneGramChannelMatrix}
    \bH \bH^\Her = \bC + \bD \bbthet \bbthet^\Her \bD^\Her \quad \in \cmplx{K \times K}.
\end{equation}
\section{Dirty Paper Coding}\label{sec:DPC}

As performance metric, we consider the sum \ac{SE}.
In the high-SNR regime a scaled identity transmit covariance matrix in the dual uplink is optimal for \ac{DPC} (see \cite{ZFDPCCompare}).
Together with \eqref{eq:RankOneGramChannelMatrix} and $\bar{p}={P_{\text{Tx}}}/K$, where $P_{\text{Tx}}$ is the transmit power, the \ac{SE} is given by 
\begin{equation}\label{eq:RateDPC}
    \begin{aligned}
        &{\text{SE}}_{\text{DPC}}  = \log_2\det(\eye + \bar{p}\bH \bH^\Her)\\
           &=\log_2\det( \eye +\bC \bar{p}) + \log_2\left( 1+ \bbthet^\Her  \bD^\Her\left(\eye/\bar{p} +\bC \right)^{\inv}  \bD \bbthet\right).\\
    \end{aligned}
\end{equation}
The asymptotic expression for $\bar{p} \rightarrow \infty$ is therefore
\begin{equation}\label{eq:RateCFullRank}
    \overline{\text{SE}}_{\text{DPC}} = \log_2\det(\bC \bar{p}) + \log_2\left( \bbthet^\Her \bD^\Her \bC^{\inv} \bD \bbthet\right) 
\end{equation}
in case $\bC$ is full rank.
If $\NB = K$ one eigenvalue of $\bC$ is exactly zero (see next section) and we obtain the asymptotic expression ($\bar{p} \rightarrow \infty$)
\begin{equation}\label{eq:RateOneEigZero}
    \overline{\text{SE}}_{\text{DPC}} = \summe{k=1}{K-1}\log_2( \lambda_k \bar{p}) + \log_2\left( \bar{p}\;\abs{\bu_K^\Her \bD \bbthet}^2\right)
\end{equation}
where $\lambda_k$ are the eigenvalues of $\bC$ in decreasing order and $\bu_k$ are the corresponding eigenvectors.
\subsection{Optimal Phase Shift Solution}
From \eqref{eq:RateOneEigZero}, we can infer that if one eigenvalue of $\bC$ is zero, we can obtain the optimal solution of the phases by alignment as 
\begin{equation}\label{eq:OptimalThetaPhaseAlign}
    \bthet = \exp\left(\imag \,\left(\angle(\bH_\tc^\Her \bu_K)-\bm{1}\angle(\bm{b}^\Her \bH_\td^\Her \bu_K)\right)\right).
\end{equation}
\paragraph*{Quadratic System ($N_{\text{\normalfont B}} =K$) } 
The matrix $\bC$ has a zero eigenvalue if $\bm{b} \in \mathrm{range}(\bH_\td^\Her)$.
For specific scenarios, this is approximately fulfilled. 
However, in case $\NB=K$, one eigenvalue of $\bC$ is always exactly zero due to the orthogonal projector $\bm{P}_{\bm{b}}^{\perp}$.
For $\bm{b} \in \mathrm{range}(\bH_\td^\Her)$, we have
\begin{equation}\label{eq:eigvalquadraticsys}
    \bu_K = \frac{\bH_\td^{+,\He} \bm{b}}{\norm{\bH_\td^{+,\He} \bm{b}}_2}
\end{equation}
where for $\NB = K$ the pseudoinverse $\bH_\td^+$ is equal to $\bH_\td^{\inv}$.
\paragraph*{Rank Improvement Scenario}
Additionally, the matrix $\bC$ also has a zero eigenvalue if $\bH_\td$ has a zero singular value.
This happens if the direct channels of some users are colinear or when we have a rank-improvement scenario where some users are blocked and, hence, have zero direct channel.
For a rank-one \ac{BS}-\ac{RIS} channel only one of the blocked users is allocated (see \cite{LOSRISDPCLinear,Eigenvalues}) and we have 
\begin{equation}
    \bu_K = \bm{e}_l
\end{equation}
where $l$ is the user with negligible direct channel.
In this case, \eqref{eq:OptimalThetaPhaseAlign} is actually a channel gain maximization of user $l$ (see \cite{LOSRISDPCLinear}).
\paragraph*{Optimal solution for high-\ac{SNR} in case $K \ge N_{\text{\normalfont B}}$}
From the discussion above, we can directly give the optimal solution for a high-\ac{SNR} scenario where we have more users than \ac{BS} antennas.
At high-\ac{SNR}, $\NB$ users will be allocated and the solution for the transmit covariance matrix is given by $\eye \bar{p}$.
According to above, the optimal phase shifts at the \ac{RIS} are given by \eqref{eq:OptimalThetaPhaseAlign} with \eqref{eq:eigvalquadraticsys} as we have a quadratic system with $\NB=K$,
due to the user allocation.

\subsection{Phase Shift Heuristic}
For $\NB = K$, the optimal phase shift vector is given by \eqref{eq:OptimalThetaPhaseAlign} with \eqref{eq:eigvalquadraticsys}.
In case $\NB > K$, we choose
\begin{equation}\label{eq:PhaseHeuristic}
    \bbthet_{\text{opt}} = \exp\left(\imag \,\left(\angle(\bw)-\bm{1}\angle(w)\right)\right).
\end{equation}
where $\bm{\bar{w}}^{\T} = [\bw^{\T},w]$ is the eigenvector corresponding to the maximum eigenvalue of $\bD^\Her(\eye/\bar{p} +\bC )^{\inv}  \bD $ to maximize ${\text{SE}}_{\text{DPC}}$ in \eqref{eq:RateDPC}.
To avoid the computation of an eigenvector in the dimension of the reflecting elements, 
we first compute the principal eigenvector $\bm{\bar{w}}^\prime \in \mathbb{C}^K$ of the matrix $(\eye /\bar{p} +\bC )^{\inv}  \bD \bD^\Her$ and then obtain the desired eigenvector (scaled version) $\bm{\bar{w}}= \bD \bm{\bar{w}}^\prime$.
The complexity of the multiplication $\bD \bD^\Her$ is only linear in $\NRe$.
This also has to be calculated only once and for each user allocation (see Section \ref{subsec:UserAlloc}) the specific submatrix can be deduced.
It is important to note, that the above heuristic phase shift configuration is optimal in case one eigenvalue of $\bC$ is zero.
The solution can be further refined with a local algorithm (e.g. element-wise) by using \eqref{eq:PhaseHeuristic} as an initial guess.

\section{Thomlinson Harashima Precoding}
To avoid the implementation difficulties of \ac{DPC}, we opt for vector perturbation precoding (see \cite{VectorPerturbation}).
One suboptimal method is \ac{THP} which avoids the computationally complex lattice search and can be employed with simple modulo transmitters/receivers.
Specifically, the sent symbol in case of \ac{THP} can be expressed as
\begin{equation}
    \bm{x} =\bm{P} \bm{v} \in \cmplx{\NB}
\end{equation}
with the transmit filter $\bm{P}$ and the symbol 
\begin{equation}
   \bv =  \mathrm{Mod}( \bm{s} + (\eye - \bm{B}) \bv ) \in \cmplx{K}
\end{equation}
after the modulo chain, where $\bm{s} \in \cmplx{K}$ is the uniformly distributed data symbol
and $\bm{B}$ is the unit lower-triangular feedback filter.
In particular, the real and imaginary part of the data symbol $s_i$ is chosen to be uniformly distributed between -0.5 and 0.5.
We use the popular reformulation (see, e.g., \cite{THPWiener}) of the modulo operation which can be equivalently written by adding an integer vector $\bm{a}$ 
resulting in $ \bv =  \bm{s} + (\eye - \bm{B}) \bv  + \bm{a}$ and, hence, in the transmitted symbol
\begin{equation}
    \bm{x} = \bm{P}\bm{B}^{\inv}(\bm{s} + \bm{a}).
\end{equation}
The receiver also employs a modulo operator and the signal
\begin{equation}
    \begin{aligned}
        \bm{y} &= \mathrm{Mod}(\bm{F} \bm{H}\bm{x} + \bm{F} \bm{n})\\
        & = \bm{F} \bm{H}\bm{x} + \bm{F} \bm{n} +\bm{\tilde{a}} 
    \end{aligned}
\end{equation}
is received. Here, $\bm{n} \sim \gaussdist{\bm{0}}{\eye}$, $\bm{F}$ is the receive filter and $\bm{\tilde{a}} $ the integer vector corresponding to the modulo operator.
\subsection{Zero-Forcing \ac{dTHP}}
Throughout this article, we consider \ac{ZF-THP}.
Combining ZF-THP with a user allocation leads to good results also for moderate \ac{SNR} values. 
There are two possibilities for ZF-THP (see, e.g., \cite{cTHPdTHP}):
\Ac{dTHP} as well as \ac{cTHP}.
We observed that \ac{dTHP} is particularly suited for a \ac{RIS}-aided scenario and we will especially focus on this version within this article.
A comment on \ac{cTHP} is given in Section \ref{sec:cTHP}.
In this subsection, we give a short overview over \ac{THP}, specifically, ZF-\ac{THP}.
For \ac{THP} as well as other approaches like MMSE-THP see \cite{THPWiener,THPZF,THPOne,THPTwo}.
Particularly, we select the transmit filters
\begin{equation}
    \bm{B} = \diag(\bm{l})^{\inv} \bm{L} , \quad \bm{P} =  \beta \bm{Q}^\Her
\end{equation} 
where $\bm{L}$ and $\bm{Q}$ are from the LQ-decomposition $\bH = \bm{L} \bm{Q}$ and $\bm{l} = [{L}_{11},\dots, {L}_{KK}]^{\T}$.
Because $\bm{B}$ is unit lower-triangular, $ \frac{1}{{L}_{ii}} \summe{j=1}{i-1} {L}_{ij} v_{j} $ from $v_i = \text{Mod}( s_i - \frac{1}{{L}_{ii}} \summe{j=1}{i-1} {L}_{ij} v_{j} )$ only depends on $s_j, \quad j<i$ and, hence, is independent of $s_i$.
As, additionally, the real and imaginary parts of $s_i$ are uniformly distributed between $-0.5$ and $0.5$, it follows that the real and imaginary parts of $v_i$ are uniformly distributed between -0.5 and 0.5 as well.
A proof for this well-known result is given in the Appendix.
Hence, with the variance of a uniform distribution, we get
\begin{equation}
    \expct{\abs{v_i}^2} = \frac{1}{6}.
\end{equation}
Since $\bQ$ is unitary, the signal power of the sent symbol $\bm{x}$ can be written as $\expct{\norm{\bm{x}}^2} = \beta^2 \frac{K}{6}$.
Hence, we obtain the scaling  
$
    \small{\beta = \sqrt{\frac{P_{\text{Tx}} 6}{K}}}
$
\normalsize
to match the power constraint $\expct{\norm{\bm{x}}^2} \le P_{\text{Tx}}$ with equality.
Choosing
\begin{equation}
    \bm{F} = \diag^{\inv}(\bm{l})\frac{1}{\beta}
\end{equation}
as the receive filter, the received symbol is given by 
\begin{equation}
    \begin{aligned}
    \bm{y} &= \text{Mod}(\bm{F}\bH\bm{P}\bm{B}^{\inv}(\bm{s} + \bm{a}) + \bm{F}\bm{n})\\
    &= \text{Mod}\left(\bm{s} + \bm{a} + \frac{1}{\beta}  \diag^{\inv}(\bm{l})\bm{n}\right)\\
\end{aligned}
\end{equation}
where $\bm{a}$ vanishes due to the modulo operation.
We arrive with $K$ complex scalar modulo channels where for each of them, we can get for the \ac{SE} (see, e.g., \cite{ModuloChannelRate,UserSelecVP,VPRatesHighSNR}) as 
\begin{equation}\label{eq:THPSE_k}
    \text{SE}_k = -h\left( \text{Mod}\left(\sqrt{\frac{K}{P_{\text{Tx}} 6  L_{kk}^2}}  \bm{n}\right)\right)
\end{equation}
with the high-\ac{SNR} asymptote (by using $\bar{p} =  P_{\text{Tx}} /K $)
\begin{equation}\label{eq:THP_Asy_k}
    \overline{\text{SE}}_k = \log_2\left(\frac{6}{\pi e} \bar{p}  L_{kk}^2 \right).
\end{equation}
Therefore, at high \ac{SNR} the sum SE reads as
\begin{equation}\label{eq:THP_Asy}
    \overline{\text{SE}} =  \log_2 \det(\bar{p} \bH \bH^\Her) -  K\log_2\left(\frac{\pi e}{6}\right)
\end{equation}
since $\det(\bH \bH^\Her) = \abs{\det{\bm{L}}}^2$. This is exactly the sum SE of \ac{DPC} at high-\ac{SNR}, however, we additionally have the shaping loss of approximately $ K\log_2\left(\frac{\pi e}{6}\right) \approx \frac{K}{2}$ bits for THP that is non-negligible.
\subsection{Phase Shift Optimization} \label{sec:PhaseShiftOpt}
As the THP high-\ac{SNR} sum \ac{SE} is just the high-\ac{SNR} \ac{DPC} sum \ac{SE} with an additional offset, the optimization of \ac{THP} at high \ac{SNR} can be directly transferred from Section \ref{sec:DPC}.
The only difference is that $\bD^\Her\bC ^{\inv}  \bD $ [see \eqref{eq:RateCFullRank}] is considered in comparison to $\bD^\Her(\eye /\bar{p} +\bC )^{\inv}  \bD $ for $\NB > K$.
In case one eigenvalue of $\bC$ is zero, e.g., when $\NB=K$, we arrive at \eqref{eq:RateOneEigZero}.
\subsection{Decoding Order}
The high \ac{SNR} sum \ac{SE} only depends on the determinant of the Gram channel matrix and, hence, is independent of the decoding order.
However, for lower \ac{SNR} values this changes and we optimize the order based on the \ac{MSE}.
This metric was e.g. used in \cite{THPWiener} and is more tractable than the \ac{SE} in \eqref{eq:THPSE_k}.
Following \cite{THPWiener}, by defining $\bm{d} = \bm{x} + \bm{a}$ at the transmitter and $\bm{\hat{d}}= \bm{y}-\bm{a}$ at the receiver, we can give the MSE as 
\begin{equation}
    \expct{\norm{\bm{\hat{d}} - \bm{d}}^2} = \expct{\norm{\bm{F}\bm{n}}^2} = \frac{ K }{P_{\text{Tx}} 6}  \summe{k=1}{K} \frac{1}{L_{kk}^2}.
\end{equation}
To avoid a computationally complex search, the decoding order is determined by a successive approach.
As in \cite{THPWiener}, we start with the user who is decoded lastly and then successively add users according to the MSE criterion.
It is important that we avoid a joint optimization of the phase shifts and the decoding order.
Hence, for a certain user allocation, we first determine the high-\ac{SNR} phase shift according to Section \ref{sec:PhaseShiftOpt} and then optimize the order for these phases.
 
\subsection{User Allocation}\label{subsec:UserAlloc}
In case, we have more users than \ac{BS} antennas or if we only have medium \ac{SNR}, it is not optimal to allocate all users for ZF-THP.
Thus, we opt for a user allocation scheme.
Specifically, we use a greedy scheme.
This is often done in the literature for similar problems and has been applied for vector perturbation precoding in \cite{UserSelecVP}.
Accordingly, we start with the first user and succesively add users one by one in a greedy manner as long as the sum SE increases.
For the sum \ac{SE}, we use the lower bound $\max(0,\overline{\text{SE}}_k)$ from \eqref{eq:THP_Asy_k} as a metric.
Note that the phase shifts as well as the decoding order have to be optimized for each allocation.
This complexity can be significantly reduced by exploiting the eigenvalue result in \cite{Eigenvalues} as has been done in \cite{LOSZeroForc} for linear precoding.
In this case, only two allocations have to be considered which clearly reduces the complexity.
Additionally, one can follow \cite{LOSZeroForc} and use a two-norm relaxation for evaluating a user allocation by replacing  $\bbthet^\Her \bD^\Her \bC^{\inv} \bD \bbthet$ with just $\NRe \lambda_{\max}(\bD^\Her \bC^{\inv} \bD)=\NRe \lambda_{\max}( \bC^{\inv} \bD\bD^\Her)$.

\section{Centralized ZF-THP}\label{sec:cTHP}
We briefly comment on \ac{cTHP} which is the alternative to \ac{dTHP} (see, e.g., \cite{cTHPdTHP}).
Here, the receive filter $\bm{F}$ at the users is an identity matrix and, hence, the receivers just consist of a simple modulo receiver.
While this reduces the receiver complexity, for \ac{cTHP}, a joint optimization of the decoding order and the phases is important.
Additionally, the high-\ac{SNR} asymptote is not just the determinant and no simple solution is available as in case of \ac{dTHP} and \ac{DPC}.
We also observed a significantly worse performance of \ac{cTHP} and in summary it appears that it is not particularly suited for a \ac{RIS}-aided scenario.
\section{Discrete Phase Shifts}
The simplicity of the objective for non-linear methods [see \eqref{eq:RateCFullRank} and \eqref{eq:RateOneEigZero}] is especially beneficial for discrete phase shifts and we also evaluate the algorithms for a binary \ac{RIS}, i.e.,  $\theta_n \in \{-1,1\}$.
The user allocation is still determined with continous phase shifts.
However, afterwards, the argument in the logarithm in \eqref{eq:RateCFullRank} and \eqref{eq:RateOneEigZero} is then optimized by applying an element-wise algorithm where the continous phase shift is chosen as an initial guess.
Note that this initial guess is not feasible and only after the first iteration a valid discrete vector is obtained.
Additionally, we extend the algorithm in \cite{LOSZeroForc} by replacing the element-wise algorithm based on continous phase shifts by an element-wise algorithm on discrete phase shifts.

\section{Simulations}
We consider an $\NB=6$ antenna \ac{BS} at (0, 0)\,m serving $K=6$ single-antenna users which are uniformly distributed in a circle  with radius 5\,m at (75, 10)\,m.
For all simulations, 3 of the users have an extra 60\,dB pathloss of the direct channel.
Hence, we simulate a rank-improvemnt scenario where at maximum 4 users will be served when including the \ac{RIS} (see \cite{LOSRISDPCLinear}).
The position of the \ac{RIS} is at (100, 0)\,m and the noise power is given by $\sigma^2=-110\text{dBm}$ in all plots.
We assume Rayleigh fading for the direct channels, Rician fading with a Rician factor of 0\,dB for the channels between the \ac{RIS} and the users and a pure \ac{LOS} channel between the \ac{BS} and the \ac{RIS}.
This pure \ac{LOS} channel is defined by an outer product of two half-wavelength \ac{ULA} vectors with both angles given by $\frac{\pi}{2}$.
For Rayleigh/Rician fading, we assume the covariance matrices to follow a Laplacian angle density (see \cite{Laplacian}) with an \ac{ASD} of $\sigma_{\text{ASD}}$.
Additionally, we choose the logarithmic pathloss model $ L_{\text{dB}} = \alpha + \beta 10\log_{10}(d)$ where $d$ is the distance in meter. 
We consider three different sets of pathloss values for different channel strengths.
Specifically, we use $L_{\text{dB},\text{weak}} = 35.1 + 36.7\log_{10}(d/\text{m})$,  $L_{\text{dB},\text{strong}} = 37.51 + 22\log_{10}(d/\text{m})$, and $L_{\text{dB},\text{LOS}} = 30 + 22\log_{10}(d/\text{m})$.

We are considering linear precoding as well as THP without the RIS (LISAwoRIS \cite{OriginalLISA}/THPwoRIS), random phase shifts (LISARandom/THPRandom) as well as optimized versions with continous (THP/AddOne-RIS-LISA) as well as discrete phase shifts (THP-Discrete/AddOne-RIS-LISA-Discrete).
Furthermore, we use \ac{DPC}-AO \cite{MaxSumRateJour} as an upper bound with 10 initial phase shifts where the best is taken.
Additionally, we include VP-Bound, which is the vector perturbation upper bound in \cite{UserSelecVP,VPRatesHighSNR,SymbolEnergyBound} with the same user allocation and phase optimization method as in this article.
Note that all \ac{THP} results include the shaping loss [see \eqref{eq:THP_Asy}].

\begin{figure}[h!]
    \centering
    \hspace*{10pt}
    \includegraphics*{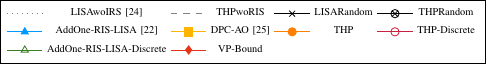}
    \subfigure[Orthogonality, $\NRe = 64$]{
    \includegraphics*{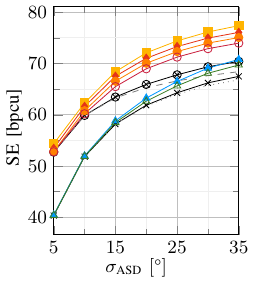}      
    \label{fig:Ortho}
    }\subfigure[RIS Elements, $\sigma_{\text{ASD}}=15\degree$]{
    \includegraphics*{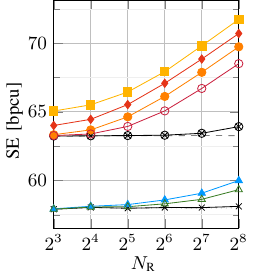}
            \label{fig:RISWeak}
    }
    \vspace*{-0.5cm}
    \caption{SE for lower impact of the \ac{RIS}}
    \label{fig:RefComp}
\end{figure}

We first analyze the scenario w.r.t. the conditioning of the channel via the \ac{ASD}.
Here, we choose the direct, the \ac{RIS}-user, as well as the \ac{BS}-\ac{RIS} channel to have an equally strong pathloss with $L_{\text{dB}\,\text{strong}}$.
In Fig. \ref{fig:Ortho}, we can see that \ac{THP} is able to handle worse conditioning better than linear precoding.
Specifically, we can see that \ac{THP} allocates a new user (via the \ac{RIS}) for a lower \ac{ASD} in comparison to linear precoding.
This has significant consequences as there are scenarios where the \ac{RIS} has strong impact in case of \ac{THP} but not for linear precoding.
This is further investigated in Fig. \ref{fig:RISWeak} based on $\sigma_{\text{ASD}} = 15\,\degree$.
Here, linear precoding has severe problems in the sense that even for a higher number of elements no or only a very slight increase in the \ac{SE} is possible.
In comparison, the \ac{SE} of \ac{THP} is significantly increased, also for a smaller number of elements.

\begin{figure}[h!]
    \centering
    \subfigure[Stronger Impact]{
        \includegraphics*{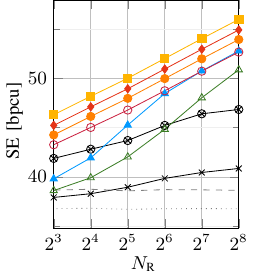}
    \label{fig:RIS}
    }\subfigure[Very Strong Impact]{

        \includegraphics*{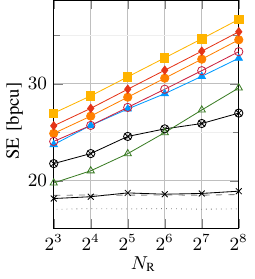}
        \label{fig:RISStrong}

    }
    \caption{SE for strong impact of the \ac{RIS} with $\sigma_{\text{ASD}} = 30\degree$}

\end{figure}

We choose now a scenario where the channels are better conditioned $\sigma_{\text{ASD}} = 30\,\degree$ and the \ac{RIS} has more impact, which is beneficial for the linear methods.
Specifically, the users are assumed to be closer to the \ac{RIS} by shifting the circle from (75, 10)\,m to (95, 10)\,m.
Additionally, we assume $L_{\text{dB, weak}}$ for the direct channels and $L_{\text{dB, LOS}}$ for the \ac{BS}-\ac{RIS} channel.
From Fig. \ref{fig:RIS} we can infer that the linear methods achieve now also a good performance.
However, a higher number of reflecting elements is needed, which is especially pronounced for discrete phase shifts.

In Fig. \ref{fig:RISStrong} we set an additional pathloss of 20\,dB for the 3 stronger users and, hence, the \ac{RIS} has an even stronger impact.
Here, we can see the fundamental limits for random and statistical phase shifts according to \cite{LOSRISDPCLinear} and linear precoding with random phase shifts is severly limited in comparison to \ac{THP} with random phase shifts.
For instantanous phase shifts on the other hand, the linear methods result in a comparable performance, even though \ac{THP} still leads to better results.
The discrete phases in case of linear precoding are still limited and there exists a major gap to the \ac{THP} methods.


\section{Conclusion}
Similar to \ac{DPC}, \ac{THP}, taking into account the shaping loss, leads to fundamental advantages in comparison to linear precoding in a \ac{RIS} assisted scenario.
Especially, when the \ac{BS}-\ac{RIS} channel is not orthogonal to the direct channel or for random/statistical phase shifts, \ac{THP} has its advantages which are even more pronounced for discrete phases.
Future work will consider non-ZF \ac{THP}, vector perturbation precoding and the important case of statistical phase shifts.

\vfill\pagebreak

\newpage
\appendix

\section{Modulo Operator \& Uniform Distribution}\label{app:UniMod}
For two independent random variables $X$ and $N$ where $X \sim \mathcal{U}(-0.5,0.5)$, the distribution
$Y = \text{Mod}(X+N)$ is again uniformly distributed between $-0.5$ and $0.5$.
To see this, we first give the distribution for $W = X+N$ which is 
\begin{equation}
    \begin{aligned}
        f_{W}(w) &= \int_{-\infty}^{\infty} f_N(\tau) f_X(w-\tau)\mathrm{d}\tau.
    \end{aligned}
\end{equation}
As $f_X(x) = 1$ for $-0.5\le x \le 0.5$ we have 
\begin{equation}
    \begin{aligned}
        f_{W}(w) &= \int_{w-0.5}^{w + 0.5} f_N(\tau) \mathrm{d}\tau.
    \end{aligned}
\end{equation}
The distribution of $Y = \text{Mod}(X+N) = \text{Mod}(W)$ is given by 
\begin{equation}
    f_Y(y) = 
    \begin{cases}
        \summe{k \in \mathbb{Z}}{} f_{W}(y + k) \quad & -0.5 \le y \le 0.5\\
        0 & \text{else}.
    \end{cases}
\end{equation}
And as, additionally, 
\begin{equation}
    \begin{aligned}
        \summe{k \in \mathbb{Z}}{} f_{W}(y + k) &= \summe{k \in \mathbb{Z}}{} \int_{w+k-0.5}^{w+k + 0.5} f_N(\tau) \mathrm{d}\tau\\
        &=\int_{-\infty}^{\infty} f_N(\tau) \mathrm{d}\tau =1
    \end{aligned}
\end{equation}
holds, we have $Y  \sim \mathcal{U}(-0.5,0.5)$.

\bibliographystyle{IEEEtran}
\bibliography{refs}
\end{document}